\documentclass[runningheads]{llncs}
\usepackage[T1]{fontenc}
\usepackage{graphicx,verbatim}
\usepackage{capt-of}
\usepackage{booktabs}
\usepackage{varwidth}
\usepackage{acro}
\usepackage{amsmath}
\usepackage{amssymb}
\usepackage[linesnumbered,ruled,vlined]{algorithm2e}
\usepackage{siunitx}
\usepackage{hyperref}
\hypersetup{
        colorlinks=true,
        linkcolor=blue,
        filecolor=magenta,
}
\usepackage{url}
\usepackage[table,dvipsnames]{xcolor}
\SetNlSty{textbf}{\color{black}}{}

\SetCommentSty{mycommentfont}
\newcommand\colornew{black}
\SetKwComment{Comment}{\# }{ }
\SetKwInput{KwInput}{Input}
\SetKwInput{KwOutput}{Output}
\DeclareAcronym{mse}{
 short=MSE,
 long=mean squared error,
}

\DeclareAcronym{mae}{
 short=MAE,
 long=mean absolute error,
}

\DeclareAcronym{mri}{
 short=MRI,
 long=magnetic resonance imaging,
}

\DeclareAcronym{ct}{
 short=CT,
 long=computed tomography,
}

\DeclareAcronym{nca}{
 short=NCA,
 long=neural cellular automata,
}

\DeclareAcronym{ae}{
 short=AE,
 long=autoencoder,
}

\DeclareAcronym{dce}{
 short=DCE,
 long=dynamic contrast-enhanced,
}

\DeclareAcronym{dwi}{
 short=DWI,
 long=diffusion weighted imaging,
}

\DeclareAcronym{auroc}{
 short=AUROC,
 long=area under the receiver operating characteristics curve,
}

\DeclareAcronym{nfs}{
 short=NFS,
 long=non-fat saturated,
}

\DeclareAcronym{fs}{
 short=FS,
 long=fat saturated,
}

\DeclareAcronym{mdl}{
 short=TeNCA,
 long=Temporal Neural Cellular Automata
}

\DeclareAcronym{gan}{
 short=GAN,
 long=generative adversarial network,
}

\DeclareAcronym{ldm}{
 short=LDM,
 long=latent diffusion model,
}

\DeclareAcronym{sota}{
 short=SOTA,
 long=state-of-the-art,
}

\DeclareAcronym{cnn}{
 short=CNN,
 long=convolutional neural network,
}

\DeclareAcronym{mlp}{
 short=MLP,
 long=multilayer perceptron,
}

\DeclareAcronym{lpips}{
 short=LPIPS,
 long=learned perceptual image patch similarity
}

\DeclareAcronym{ssim}{
 short=SSIM,
 long=structural similarity index measure
}

\DeclareAcronym{ms-ssim}{
 short=MS-SSIM,
 long=multi-scale SSIM,
}

\DeclareAcronym{psnr}{
 short=PSNR,
 long=peak signal-to-noise ratio
}

\DeclareAcronym{fid}{
 short=FID,
 long=Fréchet inception distance,
}

\DeclareAcronym{frd}{
 short=FRD,
 long= Fréchet radiomics distance
}

\newcommand{\citet}[2]{\textit{#1} \cite{#2}}
\begin{document}
\title{Temporal Neural Cellular Automata: %
Application to modeling of contrast enhancement in breast MRI}
\author{
  Daniel M. Lang\inst{1,2} \and
  Richard Osuala\inst{3} \and
  Veronika Spieker\inst{1,2} \and
  Karim Lekadir\inst{3,4} \and
  Rickmer Braren\inst{5,6} \and
  Julia A. Schnabel\inst{1,2,7}
}
\authorrunning{D.M. Lang et al.}
\titlerunning{Temporal Neural Cellular Automata}
\institute{
  Institute of Machine Learning in Biomedical Imaging, Helmholtz Munich, Germany
  \email{lang@helmholtz-munich.de} \and
  School of Computation, Information and Technology, Technical University of Munich, Germany \and
  Departament de Matematiques i Informatica, Universitat de Barcelona, Spain \and
  Institució Catalana de Recerca i Estudis Avan\c{c}ats (ICREA), Barcelona, Spain \and
  Institute for Diagnostic and Interventional Radiology, School of Medicine \& Health, Klinkum Rechts der Isar, Technical University of Munich, Germany \and
  German Cancer Consortium (DKTK), Partner Site Munich, Germany \and
  School of Biomedical Engineering \& Imaging Sciences, King's College London, UK
}

\newcommand*{\repo}{https://github.com/LangDaniel/TeNCA}
    
\maketitle              
%
\begin{abstract}
Synthetic contrast enhancement
offers fast image acquisition and eliminates the need for intravenous injection of contrast agent.
This is particularly beneficial for breast imaging,
where long acquisition times and high cost are significantly
limiting the applicability of \ac{mri} as a widespread screening modality.
Recent studies have demonstrated the feasibility of synthetic contrast generation.
However, current \ac{sota} methods lack sufficient measures for consistent temporal evolution.
\Ac{nca} offer a robust and lightweight architecture to model
evolving patterns between neighboring cells or pixels.
In this work we introduce \acs{mdl} (\acl{mdl}), which extends and further refines \acp{nca}
to effectively model temporally sparse, non-uniformly sampled imaging data.
To achieve this, we advance the training strategy by enabling adaptive loss computation and define the 
iterative nature of the method to resemble a physical progression in time.
This conditions the model to learn a physiologically plausible evolution of contrast enhancement.
We rigorously train and test \ac{mdl} on a diverse breast \ac{mri} dataset and demonstrate
its effectiveness, surpassing the performance of existing methods in generation of images that
align with ground truth post-contrast sequences.
Code: \url{\repo}
\end{abstract}
\section{Introduction}
\Acl{dce} - \acl{mri} (\acs{dce}-\acs{mri}) is the most sensitive modality for breast cancer
detection, outperforming conventional imaging with mammography, digital breast tomosynthesis and ultrasound
\cite{leithner2019abbreviated}.
The method images changes in tissue enhancement over time. 
To achieve this, multiple \ac{mri} sequences are acquired after contrast injection.
While currently reserved for supplemental screening of high-risk patients,
a growing body of evidence suggests that patients with lower risk profiles may also benefit from its use. 
However, wide adoption of \ac{dce}-\ac{mri} for breast cancer screening is
hindered by its high costs and lengthy acquisition times \cite{gao2020magnetic}.
To address this limitations, \citet{Kuhl et al.}{kuhl2014abbreviated} developed an abbreviated imaging protocol
that uses only one post-contrast image.
Nevertheless, this approach comes at the cost of losing time-resolved contrast kinetics,
which enhance specificity and enable malignancy assessment.
Ideally, a breast MRI protocol should strike a balance between high spatial resolution
and high temporal resolution, allowing for optimal diagnostic performance 
\cite{leithner2019abbreviated}.

Recent studies have shown the potential of deep learning models to predict contrast uptake from unenhanced acquisitions.
For example, \citet{Schreiter et al.}{schreiter2024virtual} developed a U-Net architecture that predicts T1-weighted
subtraction images from T1, T2, and \ac{dwi}. 
Additionally, \citet{Osuala et al.}{osuala2024towards} explored the use of \acp{ldm} to model
contrast uptake on T1-weighted breast images, conditioning on acquisition time and supplementary
imaging information using a ControlNet \cite{zhang2023adding}. 
Furthermore, the capabilities of \acp{gan} have also been investigated \cite{osuala2024simulating,kim2022tumor,muller2023using}.

\Acf{nca} are a class of models that simulate the communication and progression of cells living on a grid,
which can be effectively represented by \acp{cnn} \cite{gilpin2019cellular}.
The growing \ac{nca} variant \cite{mordvintsev2020growing} is designed to iteratively model the evolution of complex patterns.
In the medical domain, \citet{Manzanera et al.}{manzanera2021patient} extended the architecture to
simulate nodule growth in lung cancer \ac{ct}.
Additionally, \citet{Kalkhof et al.}{kalkhof2023med,kalkhof2023m3d} developed \acp{nca} for segmentation,
while \citet{Deutges et al.}{deutges2024neural} extended the architecture to classification tasks.
Furthermore, \acp{nca} have also been merged with diffusion models \cite{elbatel2024organism,mittal2025medsegdiffnca}
and applied for image registration \cite{ranem2024nca}.
The ability of \ac{nca} to model temporal textures has been investigated by
\citet{Pajouheshgar et al.}{pajouheshgar2023dynca}.
They developed a model for dynamic texture synthesis on real-world temporally-dense video data.
Moreover, \citet{Richardson et al.}{richardson2024learning} designed a nested \ac{nca}
architecture to learn spatio-temporal patterns on
artificially generated datasets featuring a uniform temporal spacing.

Typically, \ac{mri} scan times are within the order of several minutes, with the exact duration dependent on the
specific acquisition protocol.
However, the uptake and washout of contrast agent is a dynamic process that evolves continuously over time.
Therefore, a model designed to predict dynamic contrast enhancement must be capable of learning from
temporally sparse data while ensuring a continuous evolution over time.
\Acp{nca} are inherently suited for this task due to their iterative nature, which can be leveraged
to guarantee a continuous progression. 
However, this feature of \ac{nca} is usually not made use of, with the iterative process being viewed solely as
a means to reach a static output state after a fixed number of update steps.
In this work, we introduce \acs{mdl} (\acl{mdl}), a novel approach to model temporally consistent
evolution over time.
\Ac{mdl} extends the capabilities of \acp{nca} to effectively learn from temporally sparse, non-uniformly sampled
imaging data and capitalize on their iterative nature. 
To achieve this, we advance the training strategy of our model to be able to adaptively
condition intermediate states.
Furthermore, we define the update step to reflect physical progression of time,
enabling the model to simulate the continuous process of contrast enhancement.
We evaluate the performance of \ac{mdl} and compare it to two reference methods,
surpassing  current \ac{sota} performance in generation of images
that align with ground truth \ac{dce}-\ac{mri}.
Furthermore, we prove superior performance of \ac{mdl} with respect to temporal stability and
sequential consistency.
Our key contributions are as follows:
\begin{itemize}
  \item We introduce \ac{mdl}, a novel \acl{nca} based approach enabling training on temporally sparse,
    non-uniformly sampled imaging data.
  \item We adapt \ac{mdl} to model contrast enhancement on breast \ac{mri} and rigorously train and test it on a diverse dataset, involving different subcohorts and imaging protocols, with a large variety of acquisition times.
  \item We evaluate \ac{mdl} in comparison to two reference methods,
    improving current \ac{sota} performance in terms of image generation that stays close to ground truth
    post-contrast acquisitions and prove \acp{mdl} superiority in learning temporal patterns.
\end{itemize}
\section{Background: Neural Cellular Automata}
\label{sec:background}
\Acp{nca} are designed to learn update rules that allow transformation of an initial state $S_0$
into a final state $S_{\text{fin}}$, with the updates being iterative applied via \cite{pajouheshgar2023dynca}
\begin{equation}
  S_{t+1} = \mathcal{F} \left( S_t \right) = S_t + \frac{\partial S}{\partial t} \Delta t .
\end{equation}
The transition function $\mathcal{F}$ consists of a \textit{perception} and a \textit{update} part,
that can be represented utilizing a neural network \cite{gilpin2019cellular}.
The global state $S \in \mathbb{R}^{h \times w \times d}$ represents a grid of cells $s_{ij} \in  \mathbb{R}^d$
that can communicate with each other.

During the perception stage, each cell gathers information from its neighbors to form the perception vector
$z_{ij} \in \mathbb{R}^{n \cdot d}$, where $n$ represents the number of possible communication pathways.
Typically, two pathways between nearest neighbors are employed, which can be modeled using
learnable convolutional kernels \cite{deutges2024neural}.
This is combined with an identity kernel representing the cell's own state, resulting in $n=3$.
However, techniques that enable global communication can also be applied \cite{kalkhof2023med,pajouheshgar2023dynca}.
The global state $S$ is divided into two parts: visible and hidden.
\sloppy The visible part ${S_{\text{vis}} = \{s_{ijk}: i \in \{1, ..., h \}, j \in \{1, ..., w\}, k \in \{1, ..., c\}\}}$
is initialized with a seed or image of dimensionality $\mathbb{R}^{h \times w \times c}$
and the remaining hidden part stores information about cell communication.
The update part of the transition function $\mathcal{F}$ can be modeled by a \ac{mlp} via
\begin{equation}
  \frac{\partial s_{ij}}{\partial t} = \text{MLP} \left( z_{ij} \right) \odot M,
\end{equation}
where $M$ denotes a random binary variable, introduced for stochasticity \cite{pajouheshgar2023dynca}.

During training, the model weights are optimized to ensure that the iterative updates converge
to a visual part of the final state, which reflects a static target image.
Typically, the number of iterative update steps $N_{\text{steps}}$ is defined as a
hyperparameter that is empirically selected.
\section{Method: \ac{nca} for temporally sparse representations}
\label{sec:methods}
\begin{figure}[ht]
  \centering
  \includegraphics[width=0.98\textwidth]{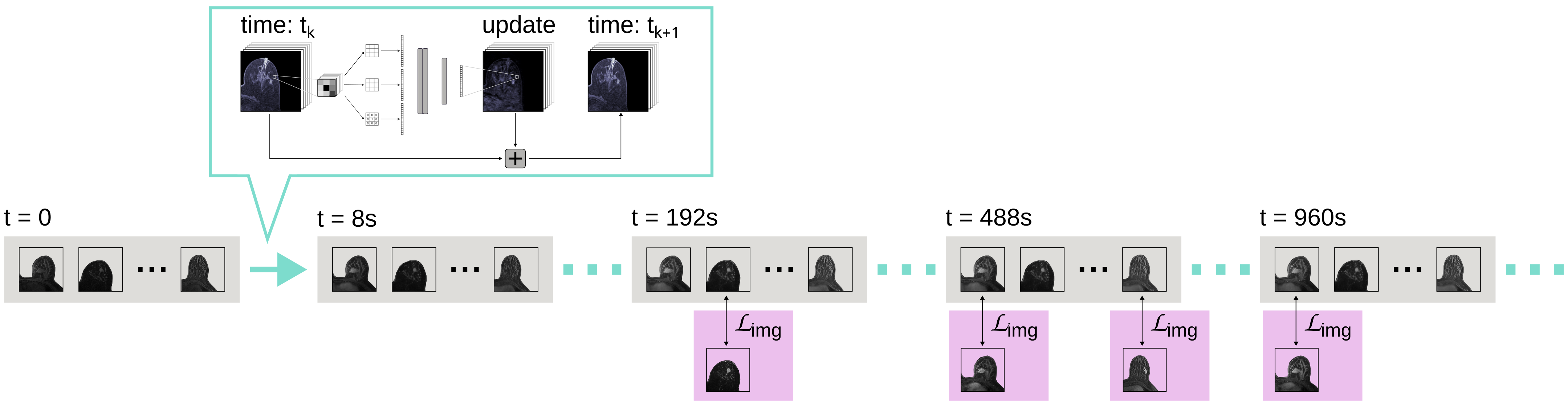}
  \caption{%
    Overview of \ac{mdl}. For each step, our {\color{teal} \ac{nca} backbone}
    transitions the images gradually
    to reflect the next time point. {\color{RedViolet}During Training},
    intermediate states are conditioned at
    all time points with a given ground truth \ac{dce}-\ac{mri} available.
  }
  \label{fig:overview}
\end{figure}

Predicting contrast uptake requires a more flexible approach than the standard \ac{nca} training procedure,
as it involves modeling a varying amount of post-contrast sequences at different time points.
To address this challenge, we extend and further develop the \ac{nca} architecture
to enable sequential loss computation.
Additionally, we define the update stage to reflect a physical progression in time,
allowing us to adaptively condition the model precisely at time points for which a ground truth
post-contrast sequence is available, see Figure \ref{fig:overview}.

Let $\{x, \{\{y_1, ..., y_k\},\{t_1, ..., t_k\}\}\}$ denote a pair of a pre-contrast image $x$ and its corresponding
post-contrast sequences $y_i$ acquired at times $t_i$ after contrast injection.
We initialize the visible part of the state $S$ with the pre-contrast image,
i.e. ${S^{\text{vis}}_0 = x \in \mathbb{R}^{h \times w \times 1}}$, while the hidden part is zero initialized.
Our goal is to have $S_{\text{vis}}$ gradually transition from a pre-contrast to a post-contrast state,
while ensuring that intermediate states also take physiologically meaningful post-contrast states.
To achieve this, we define the update step to reflect a progression in time $\Delta t$,
and require $S_{\text{vis}}$ to approximate $y_i$ after $t_i / \Delta t$ updates.
Specifically, for all update steps $\{t_i / \Delta t: i \in \{1, \dots, k\}\}$ we compute the loss between
${\hat{y}_i = S_{t_i / \Delta t}^{\text{vis}}}$ and $y_i$.
The overall loss is then given by
\begin{equation}
  \mathcal{L}
    = \sum_{j=0}^{m} \sum_{i=0}^{k_j} \mathcal{L}_{\text{img}} \left(y^j_i, S_{t_i / \Delta t}^{\text{vis}} \right)
    = \sum_{j=0}^{m} \sum_{i=0}^{k_j} \mathcal{L}_{\text{img}} \left(y^j_i, \hat{y}^j_i \right),
\end{equation}
with $k_j$ depicting the number of post-contrast sequences for patient $j$, and $m$ the number of cases
involved, i.e. given in the (mini)batch.
The loss $\mathcal{L}_{\text{img}}$ depicts a standard pixel/image based loss, e.g. \ac{mse}.
A detailed training strategy of \ac{mdl} is given in Algorithm \ref{alg:tenca}.

{
  \centering
  \begin{algorithm}[ht]\scriptsize
  \caption{Training strategy for \ac{mdl}.
  }\label{alg:tenca}
  \KwInput{
    $\mathcal{D}_{\text{train}}$: training set with pairs $\{x, \{y_1, ..., y_k\},\{t_1, ..., t_k\}\}$,
    $\mathcal{F}$: NCA transition function, $S$: NCA state, {$N_{steps}:$ update steps},
    $\Delta t$ : time-delta, $m$: batch size
  }
  \For{number of training epochs}{
    \For {$\{x^{j}, \{\{y_1^{j}, \dots, y_{k_j}^{j} \}, \{t_1^j, ..., t_{k_j}^j\}\}\}_{j=0}^{m}$ in $\mathcal{D}_{\text{train}}$}{
        \For {$j$ in $0, \dots, m$}{
      $S^{\text{vis}}_{0} \gets x^j$\\ 
      \begingroup
      \color{\colornew}
      $t \gets 0$\\
      \endgroup
      \For{l in $0, \dots, N_{steps}$}{
        \begingroup
        \color{\colornew}
        $t \gets t + \Delta t$\\ 
        \endgroup
        $S_{l+1} \gets \mathcal{F}(S_{l})$\\
        \color{\colornew}
        \For{$i$ in $1, \dots, k_j$}{
          \If{$t$ equals $t_i^j$}{
            $\hat{y}^{j}_{i} \gets S^{\text{vis}}_{l+1}$
        }
        }
      }
    }
    }
    $\mathcal{L} \gets \sum_{j=0}^{m} {\color{\colornew}\sum_{i=0}^{k_j}} \mathcal{L}_{img} \left( y_i^j, \hat{y}_i^j \right)$\\
    \textit{perform back-propagation and optimize weights of $\mathcal{F}$}
  }
  \end{algorithm}
}

By training the model in this manner, we constrain it to learn a smooth and continuous
transition of ${S}_{\text{vis}}^0=x$ into
a final state ${S}_{\text{vis}}^{N}$, while being conditioned to physiologically meaningful intermediate states
as reflected in the training data. 

\section{Experiments and Results}
Unlike previous studies \cite{muller2023using,osuala2024towards,schreiter2024virtual},
we train and evaluate our model on a diverse dataset comprising images from multiple subcohorts,
each with distinct imaging protocols.
This diversity presents a unique challenge, as the number and timing of \ac{dce} acquisitions can
vary substantially between protocols.
For instance, one clinical center might capture only two post-contrast sequences shortly after injection,
whereas another may employ five sequences to also illustrate contrast washout. 
To provide a comprehensive comparison, we test our approach against two \ac{sota} methods: 
a  U-Net model and a \acl{ldm}.

\subsection{Dataset}
For all experiments, we utilize the public MAMA-MIA dataset \cite{garrucho2024mama}
(License \href{https://creativecommons.org/licenses/by/3.0/}{CC BY 3.0} \&
\href{https://creativecommons.org/licenses/by-nc/4.0/}{CC BY-NC 4.0}
),
which comprises T1-weighted fat-saturated breast \ac{dce}-\ac{mri} scans.
We adhere to the provided training-test split, which consists of 300 test cases.
To augment the training cohort, we incorporate additional T1-weighted fat-saturated cases from
the Duke-Breast-Cancer-MRI dataset \cite{saha2018machine}
(License \href{https://creativecommons.org/licenses/by-nc/4.0/}{CC BY-NC 4.0})
that are not part of MAMA-MIA.
From this combined dataset, we randomly select 200 patients for validation,
resulting in a training set of 1604 cases.
For analysis, we consider a maximum of five post-contrast sequences and a acquisition time of up to 1024 seconds.
All images are resampled to a uniform voxel spacing of $\SI{1}{\mm}$ and intensity values are linearly
rescaled between zero and one based on the $0.02$ and $99.98$ percentiles of the respective pre-contrast image.
Additionally, images are cropped to patches of size $168 \times 168$ following the basic procedure established
by \citet{Osuala et al.}{osuala2024towards}.

\subsection{Implementation}

\noindent
\textbf{U-Net} For the implementation of the U-Net, we follow the structure of MCO-Net
\cite{schreiter2024virtual}, which models sequential post-contrast sequences through different output channels.
The method depicts a smaller variant of the standard U-Net architecture \cite{ronneberger2015u}.
However, since MCO-Net was trained on five input sequences, including T2 and \acl{dwi}, 
we perform an empirical grid search to optimize the hyperparameter.
We find that for our case the standard U-Net structure, combined with batch normalization layers,
trained on \ac{mae}, yields the best results.
The code for our U-Net model is available for reproducibility\footnotemark[1].
\footnotetext[1]{\url{\repo}}
\\

\noindent
\textbf{CC-Net} We employ the latent diffusion model-based architecture proposed by \citet{Osuala et al.}{osuala2024towards}
and adapt it to our dataset.
Specifically, we leverage the $\text{CC-Net}_{Any}$ model and retrain both the latent diffusion model
and the ControlNet architecture for 100 epochs with the given hyperparameter settings.
For encoding and decoding, we utilize the \textit{2-1-base} stable diffusion autoencoder.
\\

\noindent
\textbf{\ac{mdl}} In the perception state, \ac{mdl} utilizes two learnable kernels of size
$3 \times 3$ for communication between neighboring cells in combination with a kernel retrieving the cells own state.
We pad the input images, featuring one color channel, to a channel size of 24,
and set the temporal resolution to $\Delta t = 8$ seconds.
For the update stage, a two layered \ac{mlp} exhibiting a hidden size of 128
with the first layer using ReLU activation is employed.
We train \ac{mdl} with \ac{mse} as the image loss and perform empirical
hyperparameter optimization.
We make the code for \ac{mdl} available for reproducibility\footnotemark[1]. 

\subsection{Evaluation Metrics}
Image evaluation metrics include \ac{lpips} \cite{zhang2018unreasonable},
the \ac{ssim} and \ac{ms-ssim} \cite{wang2003multiscale}, as well as \ac{psnr}.
Distribution measures involve \ac{fid} \cite{heusel2017gans} and \ac{frd} \cite{osuala2024towards}.
As models were trained to optimize different losses, i.e. \ac{mse} and \ac{mae},
and are, therefore, likely biased towards their respective training objective, we do not include those metrics in our analysis. 
As a lower/upper bound, we compute the difference between each post-contrast acquisition and its respective pre-contrast image,
the result of which we denote as baseline.

\subsection{Results}
\begin{figure}[ht]
  \centering
  \includegraphics[width=0.99\textwidth]{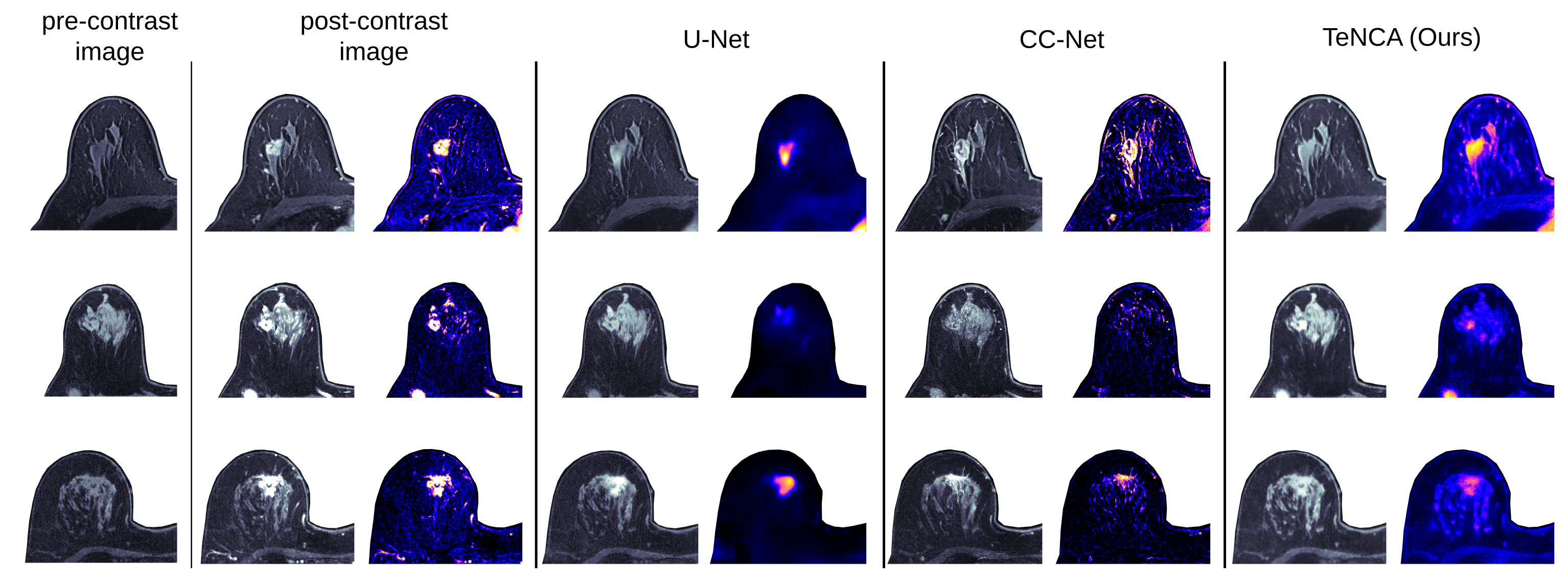}
  \caption{%
    Example test set results, involving a (predicted) post-contrast image and a subtraction
    between the pre- and the post-contrast image, highlighting contrast uptake.
    \Ac{mdl} successfully models detailed structures, outperforming the U-Net.
    Additionally, it avoids hallucination of artifacts, a limitation evident in the first example of CC-Net.
    }
  \label{fig:results}
\end{figure}

The overall model performance on the test set, calculated as the mean across all post-contrast phases,
is presented in Table \ref{tab:results}.
Qualitative results are visualized in Figure \ref{fig:results}.
\newcommand{\ic}{\cellcolor{teal!25}}
\newcommand{\dc}{\cellcolor{RedViolet!25}}
\begin{table}
\centering
\caption{%
  {Comparison of \color{teal}Image metrics} and {\color{RedViolet}distribution measures} on the test set,
  alongside model parameter counts. Notably,  \ac{mdl} excels in image metrics while maintaining competitive
  distribution measure values. CC-Net achieves higher values for distribution measures,
  but its image metric performance suggest a tendency to hallucinate image parts.
  \ac{mdl} requires significantly less parameters.
}
\begin{tabular}{p{0.17\textwidth}p{0.1\textwidth}p{0.08\textwidth}p{0.13\textwidth}p{0.15\textwidth}p{0.075\textwidth}p{0.075\textwidth}p{0.12\textwidth}}
  \toprule
  Method              &\ic LPIPS$\downarrow$  &\ic SSIM$\uparrow$   &\ic MS-SSIM$\uparrow$  &\ic PSNR [dB]$\uparrow$  &\dc FID$\downarrow$  &\dc FRD$\downarrow$ & param.$\downarrow$   \\
  \midrule                                                  
  baseline            & 0.13                  & 0.86                & 0.88                  & 29.24                   & 30.20               & 154.20          & -\\
  U-Net               & 0.13                  & \underline{0.88}    & \underline{0.92}      & \underline{31.93}       & 32.00               & 50.96           & $31 \cdot 10^{6}$ \\
  CC-Net              & 0.14                  & 0.80                & 0.80                  & 30.19                   & \textbf{21.28}      & \textbf{20.00}  & $12 \cdot 10^{8}$\\                                                   
  \midrule                                                  
  TeNCA(ours)        & \textbf{0.12}         & \textbf{0.89}       & \textbf{0.93}         &\textbf{32.26}           & \underline{27.83}   &\underline{48.68}& $\mathbf{13 \cdot 10^{3}}$\\
  \bottomrule
\end{tabular}
\label{tab:results}
\end{table}
Notably, \ac{mdl} achieves the highest overall performance,
surpassing all other methods in terms of image-level metrics.
However, CC-Net outperforms \ac{mdl} in distribution similarity metrics,
specifically FID and FRD, which assess the similarity between the set of 
generated images and the ground truth \ac{dce} dataset. 
This suggest that CC-Net is capable of producing more realistic-looking images.
Nevertheless, CC-Net's performance in image-level metrics reveals a significant limitation: 
at pixel level, the generated images deviate substantially from the ground truth post-contrast sequences,
as indicated by \ac{lpips} and (MS-)\ac{ssim} values below the baseline.
This implies that CC-Net is prone to hallucinating parts in the images,
a known issue with diffusion models \cite{aithal2025understanding}.
An example of this can be seen in the first row of Figure \ref{fig:results},
CC-Net generates a realistic-looking example that, however, fails to reflect the actual post-contrast sequence.
In contrast, the U-Net architecture performs the worst, with image metrics lower than \ac{mdl}
and the lowest results for distribution measures.
The qualitative examples in Figure \ref{fig:results} suggest a segmentation-like behavior,
which is consistent with the task for which the U-Net architecture was initially designed
\cite{ronneberger2015u}.
As depicted in Table \ref{tab:results}, \ac{mdl} requires significantly less parameters than other methods.\\

\begin{figure}[ht]
  \centering
  \includegraphics[width=0.99\textwidth]{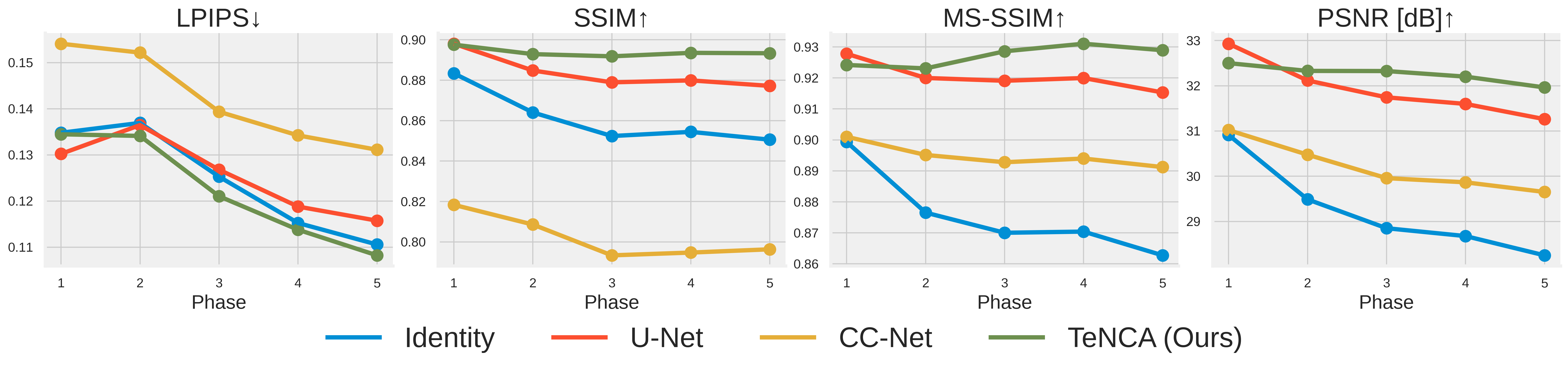}
  \caption{%
    Mean image metric values for the test set across all post-contrast phases.
    \Ac{mdl} maintains consistent performance throughout all phases,
    while other methods exhibit a noticeable decline in later phases.
  }
  \label{fig:results_seq}
\end{figure}

\noindent
\textbf{Temporal stability} Figure \ref{fig:results_seq} illustrates the image metric performance
calculated individually for each post-contrast phase.
\Ac{mdl} consistently features superior temporal ability. 
Other methods exhibit a decline in performance as time progresses/for later phases, with
the U-Net mostly achieving its best results for the first phase.
In contrast, \ac{mdl} achieves stable temporal performance with \ac{ms-ssim} even improving
for later phases.\\

\noindent
\textbf{Sequential consistency} Example videos illustrating the temporal evolution of
contrast enhancement are provided online\footnotemark[2].
\footnotetext[2]{\url{https://langdaniel.github.io/TeNCA/}}
Notably, \ac{mdl} exhibits the best sequential consistency,
characterized by a continuous evolution.
In contrast, CC-Net's output changes significantly between consecutive frames, highlighting
the inevitable need for temporal stability measures to be taken into account.
Additionally, the segmentation-like behavior of the U-Net model is evident,
with its output remaining relatively static.
%
\section{Discussion and Conclusion}
\label{sec:discussion}
This paper introduces \ac{mdl}, a novel approach that enhances the training procedure
of \acl{nca} to effectively model temporally sparse, non-uniformly sampled imaging data.
We train \ac{mdl} to predict contrast enhancement on breast \ac{mri}
and comprehensively evaluate its performance on a challenging dataset including diverse sub-cohorts
with varying imaging protocols and acquisition times.
Our results demonstrate the superiority of \ac{mdl} over two existing methods,
surpassing current \ac{sota} performance in generation of images that align with ground truth
post-contrast sequences.
Furthermore, we prove \ac{mdl}'s strong temporal capabilities, which performs consistent
over all post-contrast phases and evolves its output continuously over time.
Notably, \ac{mdl} is less susceptible to hallucinations,
an issue with diffusion based contrast prediction,
which poses a significant concern in the medical domain where
algorithm reliability is paramount for clinical applicability \cite{lekadir2025future}. 
Additionally, \ac{mdl} requires substantially fewer parameters than other methods, making in easily deployable,
even in resource-constrained settings.

The strong performance of \ac{mdl} motivates us to further enhance its capabilities to 3D modeling
and evaluate its clinical applicability in future work.
Its flexible training strategy also opens new opportunities for application,
e.g. in cine \ac{mri} or 4D \ac{ct}.

\begin{credits}
\subsubsection{\ackname}%
DML and JAS received funding from HELMHOLTZ IMAGING, a platform of the Helmholtz Information \& Data Science Incubator.
VS is partially supported by the Helmholtz Association under the joint research school "Munich School for Data Science (MUDS)".
This project (RO, KL) has received funding from the EU Horizon Europe and Horizon 2020 research and innovation programme under
grant agreement No 101057699 (RadioVal) and No 952103 (EuCanImage), respectively.
RO acknowledges a research stay grant from the Helmholtz Information and Data Science Academy (HIDA).

\subsubsection{\discintname}%
All authors declare that they have no conflicts of interest.
\end{credits}
%
%
\bibliographystyle{splncs04}
\bibliography{references}
\end{document}